\documentclass{article}
\usepackage{spconf,eqnarray,graphicx,textcomp}
\usepackage{booktabs, multirow}
\usepackage{amssymb} 
\usepackage{tikz} 
\usepackage{mathtools} 
\usepackage{svg}
\usepackage{amsmath}
\usepackage{hyperref}
\usepackage{caption}
\usepackage{subcaption}

\title{Improving performance of real-time full-band blind packet-loss concealment with predictive network}

\name{Viet-Anh Nguyen$^{1}$, Anh H. T. Nguyen$^{1}$, and Andy W. H. Khong$^{2}$}
\address{$^{1}$Crystalsound Team, NamiTech JSC, Ho Chi Minh City, Vietnam\\
$^{2}$Nanyang Technological University, Singapore\\
\texttt{\{vietanh.nguyen, anh.nguyen\}@namitech.io, andykhong@ntu.edu.sg}
}

\begin{document}
%\ninept
%

\maketitle

\begin{abstract}
Packet loss concealment (PLC) is a tool for enhancing speech degradation caused by poor network conditions or underflow/overflow in audio processing pipelines. We propose a real-time recurrent method that leverages previous outputs to mitigate artefact of lost packets without the prior knowledge of loss mask. The proposed full-band recurrent network~(FRN) model operates at 48~kHz, which is suitable for high-quality telecommunication applications. Experiment results highlight the superiority of FRN over an offline non-causal baseline and a top performer in a recent PLC challenge.
\end{abstract}

\vspace{-0.25cm}
\section{Introduction}
\vspace{-0.15cm}
\label{sec:intro}
The advent of technology for remote communication devices coupled with the recent pandemic has resulted in the rise in demand for voice over Internet Protocol (VoIP) in many communication systems. Besides the need to deal with network echoes~\cite{Khong2007ALD, zhang22t_interspeech, andy}, packet loss may occur leading to artefacts and distortion at the receiver end. Packet loss concealment (PLC) aims to mitigate these effects by filling in the gaps due to lost packets.

There are two types of PLC algorithms\textemdash informed and blind PLC~\cite{Kegler_2020}. The former requires prior information pertaining to which audio packets have been lost during transmission. In this regard, conventional methods employ linear or statistical models to conceal unavailable audio packets~\cite{phmm2, lpc}. Deep learning methods, on the other hand, can conceal faulty chunks from its representation in the time-~\cite{crn} or time-frequency domain~\cite{rtime, Kegler_2020} without the need of feature engineering. In the recent PLC challenge~\cite{Diener2022INTERSPEECH2A}, deep neural networks such as~\cite{plc1, plc2, plc3} have been successfully adopted for informed PLC, achieving promising results. 

As opposed to the informed approach, blind PLC, or audio in-painting, improves lossy signals without the need of loss traces and is suitable when packet metadata may not be readily available. In addition, these algorithms do not require the alignment of audio input with its packet stream resulting in their ability to handle arbitrary packet sizes. An end-to-end speech inpainting network has been proposed to recover losses in both time and frequency axes of the spectrogram representation~\cite{Kegler_2020}. Generative adversarial networks (GANs) have also shown promising results for this task~\cite{tfgan}. It is useful to note that most of existing PLC methods operate at 16~kHz sampling rate, which presents a limitation for full-band 48 kHz transmission. Furthermore, in-painting models~\cite{Kegler_2020, tfgan} are non-causal, rendering them unsuitable for real-time communication.
\begin{figure*}
    \centering
    \includegraphics[width=0.9\textwidth]{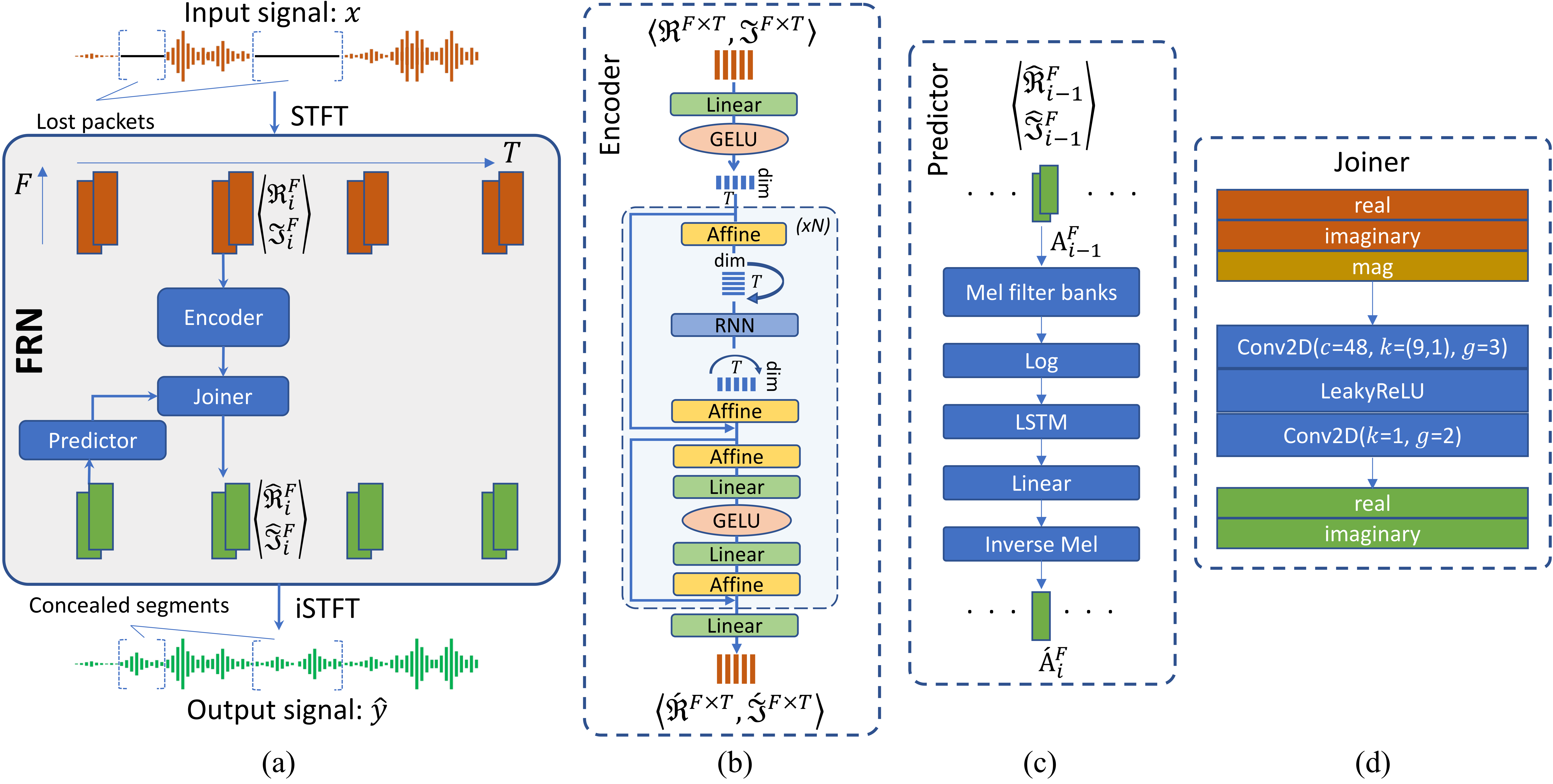}
    \caption{Full-band recurrent network (FRN) (a) that processes time-frequency frames sequentially. The architecture consists of an encoder (b) that enhances the current input, a predictor (c) that infers output from the previous step, and a joiner (d) that combines outputs of the two modules.}
    \label{fig:method}
\end{figure*}

We propose a full-band recurrent network (FRN) for blind PLC of 48 kHz speech signals. The proposed FRN algorithm is frame-causal due to the use of short-time Fourier transform (STFT) and does not require any additional information (e.g., loss mask or packet size) of the packet stream. As opposed to conventional networks~\cite{Kegler_2020, crn, tunet}, and inspired by the recurrent neural network (RNN)-transducer model~\cite{transducer} in speech recognition, our proposed FRN model includes a log Mel-scale predictor to address practical challenges posed by high sample-rate, unavailable packet-loss metadata, and causal frame-wise inference. The predictor employs the previous output to improve prediction of the current output, which is essential to achieve better concealment when several consecutive packets are lost.
% We combine features from two branches by two individual sub-modules: an encoding branch that enhances the current frame, and a predictive branch that predicts the current output from precedent outputs. The encoding branch would be beneficial when the current time-frequency frame partially contains loss packets, whereas the predictive branch helps when the content in the current frame has lost completely. We further employ data augmentation that could be necessary for the method to handle audio signals at lower sampling rates. 
It is worth noting that the employment of previous output in speech enhancement algorithms has a long history, e.g., in the celebrated decision-directed SNR estimator~\cite{reuse}. Experiment results demonstrate that FRN outperforms an informed PLC model and an offline GAN-based model despite being a blind frame-causal algorithm.
\vspace{-0.15cm}
\section{The proposed full-band recurrent network (FRN)}
\vspace{-0.1cm}
% Our proposed FRN model is inspired by the RNN-Transducer \cite{transducer}, which consists of an encoder $Enc$ and a predictor $Pred$. Since our method is a blind PLC algorithm, we assume that content in each input frame is partially or completely lost in this work. 
We assume that the content of each input chunk may be partially or completely lost in the time domain and that information pertaining to which part of the input signal has been lost is unavailable. To conceal such audio loss in online manner, we propose a time-frequency model consisting of an encoder $\mathrm{Enc}$ and a predictor $\mathrm{Pred}$ as shown in Fig.~\ref{fig:method}(a). Input signal $x$ is represented via $T$ frames and $F$ frequency components by employing STFT, resulting a complex time-frequency (TF) representation $\langle  \substack{\mathfrak{R}^{F\times T}, \mathfrak{I}^{F\times T}}\rangle ^\intercal = \operatorname{STFT}(x)$. At a time step $i$, the encoder reconstructs audio loss in the current input $\langle  \substack{\mathfrak{R}^{F}_i, \mathfrak{I}^{F}_i}\rangle ^\intercal$ while the pretrained predictor leverages previous output $\langle  \substack{\hat{\mathfrak{R}}^{F}_{i-1}, \hat{\mathfrak{I}}^{F}_{i-1}}\rangle ^\intercal$ and exploits predictable patterns in speech signals to yield further improvement. Finally, the enhanced TF frame $\langle  \substack{\hat{\mathfrak{R}}^{F}_{i}, \hat{\mathfrak{I}}^{F}_{i}}\rangle ^\intercal$ is generated via a joiner that combines the encoder output and the predictor output. 

The detailed architecture of our encoder is illustrated in Fig.~\ref{fig:method}(b). In essence, the proposed encoder models inter-frame information via an RNN for causality and models intra-frame information via a multi-layer perceptron (MLP) with fusible affine normalization layers. Such a network arrangement can be seen as an efficient hybrid between ResMLP~\cite{resmlp} and dual path RNN~\cite{dprnn}. The complex-value STFT of the input is first flattened into $\mathfrak{RI}^{2F\times T}=\langle  \mathfrak{R}^{F\times T}, \mathfrak{I}^{F\times T} \rangle$ and projected from the dimension $2F$ (real and imaginary) to a lower dimension $dim$ for computational efficiency. This projection is achieved via a linear layer followed by a Gaussian error linear unit (GELU) resulting in feature
\begin{equation}
    x^{dim \times T} = \operatorname{GELU}(W_1^{dim\times 2F}\cdot\mathfrak{RI}^{2F\times T} + b_1),
\end{equation}
where $W_1^{dim\times 2F}$ and $b_1$ are weights and bias of the linear layer, respectively. The subsequent residual module $\mathrm{Res}$ consists of $N$ blocks with each block comprising an RNN layer that models the inter-frame correlation and linear layers that model intra-frame features. Features of the residual module $\mathrm{Res}$ is then projected back to dimension of $2F$ giving the encoder output
\begin{equation}
    \left\langle\Acute{\mathfrak{R}}^{F \times T}, \Acute{\mathfrak{J}}^{F \times T}\right\rangle = W_2^{2F\times dim}\cdot \mathrm{Res}(x^{dim \times T}) + b_2,
\end{equation}
where $W_2^{2F\times dim}$ and $b_2$ are weights and bias of the projection layer, respectively. 

% \subsection{Predictor}
The architecture of the predictor $\mathrm{Pred}$ is shown in Fig.~\ref{fig:method}(c). It predicts the magnitude $\Acute{A}_i^F$ of the current frame $i$ from that of the previous complex output frame via
\begin{align}
A_{i-1}^F &= \sqrt{(\mathfrak{R}^F_{i-1})^2 + (\mathfrak{I}^F_{i-1})^2}, \\
\Acute{A}_i^F &= \mathrm{Pred}(A_{i-1}^F).
\end{align}
Since significant amount of speech energy is concentrated in the lower frequency-bin indices, the predictor is simplified and made efficient by applying a long short-term memory (LSTM) on the log Mel-scale magnitude input. Here, the LSTM models the temporal dynamics of the signal and a  subsequent linear layer projects the LSTM output back to the Mel-scale dimension. Finally, a learnable inverse Mel layer consisting of an exponential function, a linear layer, and an absolute activation transforms the Mel magnitude into STFT magnitude of $F$ dimension. To ensure stability during training and to ensure the prediction from the previous input frame, the predictor is pre-trained on the target audios of the training PLC dataset before being jointly trained with the encoder network. This pre-training step allows the model to achieve modest improvement in performance.

To combine magnitude feature from the predictive branch with the complex feature from the encoder, we employ a joiner based on the convolutional neural network (CNN). Since the two features have the same shape $F\times T$ but differ in the number of channels, CNNs serve as an effective tool for feature fusion. Features are stacked along its channel axis, forming a 3-channel input $x$ including $\Acute{\mathfrak{R}}^F_i, \Acute{\mathfrak{I}}^F_i$, and $\Acute{A}^F_i$\, which denote, respectively, the real, imaginary parts of the encoder output, and the magnitude output of the predictor.

Two causal grouped convolution layers are used to transform the three channels into two channels representing the real $\hat{\mathfrak{R}}^F_i$ and imaginary $\hat{\mathfrak{I}}^F_i$ of the final output before being transformed into waveform $\hat{y}$ by inverse STFT (iSTFT), i.e,
% \vspace{-0.1cm}
\begin{align}
 \langle \substack{\hat{\mathfrak{R}}^F_i, \hat{\mathfrak{I}}^F_i} \rangle ^ \intercal 
 &= \operatorname{Joiner}( \langle \substack{\Acute{\mathfrak{R}}^F_i, 
                            \Acute{\mathfrak{I}}^F_i, 
                            \Acute{A}^F_i} \rangle ^ \intercal), \\
\hat{y} &= \operatorname{iSTFT}(\langle \substack{\hat{\mathfrak{R}}^F_i, 
                            \hat{\mathfrak{I}}^F_i} \rangle ^ \intercal).
\end{align}
% \vspace{-0.9cm}
\subsection{Learning objective}
\vspace{-0.1cm}
We employ multi-resolution STFT loss~\cite{mulres} as the learning objective. Given a signal $s$, its compressed STFT magnitude $\mathcal{C}(s)$ is defined as
\begin{equation}
\label{stft2}
    \mathcal{C}(s) = |\operatorname{STFT}(s)|^\alpha,
\end{equation}
where $|\operatorname{STFT}(s)|$ denotes the STFT magnitude of $s$, and $\alpha$ is a compression rate that equalizes energy across frequency bands~\cite{compress}. The multi-resolution STFT loss $\ell_{\mathrm{MR}}$ between the generated signal $\hat{y}$ and its corresponding target signal $y$ over $M$ resolutions is then given by
\begin{equation}
\resizebox{7.5cm}{!}{%
    $\ell_{\mathrm{MR}}(\hat{y}, y)=\frac{1}{M} \sum_{m=1}^{M}\left(
    \frac{\|\mathcal{C}(y)-\mathcal{C}(\hat{y})\|_{\mathrm{F}}}{\|\mathcal{C}(y)\|_{\mathrm{F}}} + 
    \frac{1}{N}\|\mathcal{C}(y)-\mathcal{C}(\hat{y})\|_{1}
    \right)$,
}
\end{equation}
where the first term within the brackets corresponds to spectral convergence while the second term is for spectral magnitude loss, $\| . \|_{F}$, $\| . \|_{1}$, and $N$ are the Frobenius norm, L1 norm, and total number of frequency bins, respectively. 

\subsection{Packet loss generation}
\label{sec:plgen}
To simulate realistic packet loss sequences, a two-state Markov chain algorithm consisting of `loss' ($L$), `non-loss' ($N$), and transition probabilities are used to control the behaviour of the output sequence \cite{crn}. Defining $p_N$ and $p_L$ as the intra-state transition probabilities of state $N$ and $L$, respectively, a two-state Markov chain can be defined as a ($p_N$, $p_L$) pair. We can modify these probabilities to achieve different expected loss rates of the packet sequences.

\vspace{-0.1cm}
\section{Experiment setup and results}
\vspace{-0.1cm}
\label{sec:experiments}

\subsection{Setup}
% \subsubsection{Dataset} 
The dataset was generated from the VCTK Corpus~\cite{Veaux2017CSTRVC} by selecting three-second segments randomly from each audio file to create the target signal. The lossy input signal was created by splitting the target signal into packets with size chosen arbitrarily from the set $\{256, 512, 768, 1024, 1536\}$. A loss mask generated from the procedure in Section~\ref{sec:plgen} is then applied on the packetized signal. The dataset consists of 109 English speakers from which audio clips of the first 100 speakers are for training and the remainder for testing. Loss masks were generated, for each audio output, using one of the four Markov chains in Table~\ref{tab:mc}. Besides, we also use loss masks provided in the PLC challenge \cite{Diener2022INTERSPEECH2A} for realistic evaluation. These loss traces are collected from real calls and reflects the behaviour of lost packets in the real-world.
% \usepackage{booktabs}
% \usepackage{multirow}
\begin{table}[]
\centering
\begin{tabular}{@{}lll@{}}
\toprule
\textbf{$p_N$} & \textbf{$p_L$} & \textbf{Expected loss} \\ \midrule
0.9         & 0.1         & 10\%                   \\
0.9         & 0.5         & 16.7\%                 \\
0.5         & 0.1         & 35.7\%                 \\
0.5         & 0.5         & 50\%                   \\ \bottomrule
\end{tabular}
\caption{Transition probabilities of Markov chain models}
\label{tab:mc}
\end{table}

\begin{table*}[]
\centering
\resizebox{\textwidth}{!}{%
\begin{tabular}{@{}|l|llll|llll|llll|llll|@{}}
\toprule
\multirow{2}{*}{\textbf{Model}}                              & \multicolumn{4}{c|}{\textbf{PLCMOS}}                                                                             & \multicolumn{4}{c|}{\textbf{LSD}}                                                                                 & \multicolumn{4}{c|}{\textbf{STOI}}                                                                                         & \multicolumn{4}{c|}{\textbf{PESQ}}                                                                                         \\ \cmidrule(l){2-17} 
                                                             & \textbf{10\%}  & \textbf{20\%}  & \textbf{40\%}  & \textbf{\begin{tabular}[c]{@{}l@{}}Real\\ trace\end{tabular}} & \textbf{10\%}  & \textbf{20\%}  & \textbf{40\%}  & \textbf{\begin{tabular}[c]{@{}l@{}}Real \\ trace\end{tabular}} & \textbf{10\%}  & \textbf{20\%}  & \textbf{40\%}  & \textbf{\begin{tabular}[c]{@{}l@{}}Real \\ trace\end{tabular}} & \textbf{10\%}  & \textbf{20\%}  & \textbf{40\%}  & \textbf{\begin{tabular}[c]{@{}l@{}}Real \\ trace\end{tabular}} \\ \midrule
\begin{tabular}[c]{@{}l@{}}Input \\ (zero fill)\end{tabular} & 3.505          & 2.754          & 1.931          & 3.517                                                         & 1.156          & 1.811          & 3.267          & 1.984                                                          & 0.842          & 0.748          & 0.592          & 0.797                                                          & 1.931          & 1.393          & 1.126          & 2.484                                                          \\
tPLC                                                         & 3.417          & 2.863          & 2.417          & 3.463                                                         & 1.201          & 1.611          & 2.073          & 1.216                                                          & 0.831          & 0.776          & 0.704          & 0.838                                                          & 1.981          & 1.629          & 1.386          & 2.532                                                          \\
TFGAN                                                        & 3.902          & \textbf{3.675} & \textbf{3.019} & 3.645                                                         & 2.081          & 3.181          & 3.714          & 2.314                                                          & 0.823          & 0.751          & 0.641          & 0.760                                                          & 1.716          & 1.321          & 1.116          & 1.339                                                          \\
\begin{tabular}[c]{@{}l@{}}FRN \\ (proposed)\end{tabular}                                                         & \textbf{4.032} & 3.573          & 2.623          & \textbf{3.655}                                                & \textbf{0.783} & \textbf{1.117} & \textbf{1.682} & \textbf{0.946}                                                 & \textbf{0.920} & \textbf{0.862} & \textbf{0.746} & \textbf{0.889}                                                 & \textbf{2.336} & \textbf{1.799} & \textbf{1.422} & \textbf{2.797}                                                 \\ \bottomrule
\end{tabular}
}
\caption{Score comparison with baselines. The `10\%', `20\%', and `40\%' columns indicate test sets with simulated loss masks while the `Real trace' column indicates test set with loss traces (approx. loss rate of 10.27\%) from real calls. LSD lower is better while the other three higher is better.}
\label{tab:res}
\end{table*}
% \subsubsection{Evaluation metrics}
To evaluate performance, we used four metrics: PLCMOS \cite{Diener2022INTERSPEECH2A}, log-spectral distance (LSD), short-time objective intelligibility (STOI), and perceptual evaluation of speech quality (PESQ). While STOI, PLCMOS, and PESQ only evaluate frequency bands of up to 5~kHz, 8~kHz, and 8~kHz, respectively, LSD evaluates the entire 24~kHz frequency band of the full-band speech. PLCMOS is the average score of two independent deep learning models that predict intrusive and non-intrusive mean-opinion-score (MOS) of human raters specifically for PLC task. LSD, STOI, and PESQ, on the other hand, rely on feature engineering to capture characteristics of speech signals associated with auditory quality. Higher PLCMOS, STOI, and PESQ scores imply better quality while a lower LSD score is more ideal.
% \bigskip

% \subsubsection{Hyperparameters}
We applied 960-point STFT with window size of 20~ms and 50\% overlapping on the audio waveform, resulting in the dimension of $F=480$ for each STFT frame. While lookahead was not applied to reduce overall algorithmic latency, it can be used if performance is prioritized. For the encoder, the projection dimension $dim$ and RNN state dimension are equivalent to 384 while the hidden dimension of the subsequent linear layers and the number of layers are $N=768$ and 4, respectively. For the predictor, we used 64-bin Mel filter banks, and the LSTM consists of a single layer with a hidden size of 512. The hyperparameters of convolution layers in the joiner, including output channel $c$, kernel size $k$, and number of group $g$, are shown in Fig.~\ref{fig:method}(d). The STFT paramaters for the resolutions in our loss function is set to default values of the \textit{auraloss} library~\cite{auraloss}, and the compression ratio is set to $\alpha=0.3$. For the PLC training as well as predictor pretraining, we trained the models for 150 epochs with 90 in samples each data batch. Weights of the models are optimized by the Adam optimizer with $1\times 10^{-4}$ learning rate. 

\subsection{Results}
% \label{sec:per}
For comparison, we implemented two baselines:
\begin{itemize}
    \item \textbf{tPLCnet}~\cite{plc3}: a real-time informed PLC method that was ranked third in the recent PLC challenge. We employed the `Large' model, which is the largest version proposed by the authors.
    \item \textbf{TFGAN}~\cite{tfgan}: an offline blind PLC model based on generative adversarial networks (GANs). Since the model was designed for concealing 16~kHz audios, chunk size and discriminator linear head were increased from 2,560 to 7,680 to better adapt to 48~kHz signals.
\end{itemize}
The models were evaluated on the VCTK test set with the packet size fixed at 20~ms (i.e., 960 samples). The loss mask was generated by three Markov chains, forming three simulated test sets with expected loss rates being 10\%, 20\%, and 40\%. Another test set is `Real trace', which corresponds to the real loss traces derived from the PLC challenge dataset applied on the VCTK test set.

With reference to Table~\ref{tab:res} and compared to the informed tPLCnet method~\cite{plc3}, the proposed model achieved significantly higher scores across all benchmarks and loss rates despite the lack of loss trace information. We also note that although the offline GAN-based TFGAN model achieved higher PLCMOS scores, especially at higher loss rates such as 40\%, our model outperforms TFGAN by approximately 250\%, 12\%, and 30\% in terms of LSD, STOI, and PESQ measures, respectively. To evaluate the listening quality, we have also provided audio samples along with our source code for comparison\footnote{\url{https://github.com/Crystalsound/FRN}}.

% \usepackage{booktabs}
\begin{table}[]
\centering
\resizebox{\linewidth}{!}{%
\begin{tabular}{@{}lccccc@{}}
\toprule
\textbf{Model}                                                 & \textbf{Causal} & \textbf{Blind} & \begin{tabular}[c]{@{}c@{}}\textbf{Model} \\ \textbf{size}\end{tabular} & \begin{tabular}[c]{@{}c@{}}\textbf{Inference} \\ \textbf{time}\end{tabular} & \begin{tabular}[c]{@{}c@{}}\textbf{Real-time} \\ \textbf{factor}\end{tabular} \\ \midrule
tPLCnet                                                        & \checkmark      &       & 5.7M                                                  & 2.7 ms      & 0.27                                              \\ \midrule
\begin{tabular}[c]{@{}l@{}}TFGAN   \\ (generator)\end{tabular} &        & \checkmark     & 1.9M                                                  & N/A  & N/A                                                       \\ \midrule
\begin{tabular}[c]{@{}l@{}}FRN \\ (proposed)\end{tabular}                                                            & \checkmark      & \checkmark     & 9.1M                                                  & 4.1 ms  & 0.41                                                    \\ \bottomrule
\end{tabular}
}
\caption{Model size, inference time, and real-time factor of the algorithms. `Causal' column indicates causality of the models, and `Blind' column indicates blind PLC algorithms}
\label{tab:time}
\end{table}

We evaluated the inference time of the models via a single Intel Core-i9 3.0 GHz CPU thread and the ONNX inference engine. Since TFGAN is a non-causal model, we only evaluated the latency of our model and tPLCnet. As shown in Table~\ref{tab:time}, although our proposed model suffers from a modest 1.4~ms increase in inference time than tPLCnet, its increase in performance outweighs the higher inference time. In particular, the proposed FRN model requires 4.1 ms to process a 10 ms chunk -- 20ms windows with 50\% overlapping -- achieving a real-time factor of 0.41.

To gain insight into the impact of the predictor, we conduct experiments on the `Real trace' test set to compare FRN with the encoder only.
\begin{table}[]
\centering
\resizebox{7cm}{!}{%
\begin{tabular}{@{}lclll@{}}
\toprule
\textbf{Model}    & \multicolumn{1}{l}{\textbf{PLCMOS}} & \textbf{LSD}   & \textbf{PESQ}  & \textbf{STOI}  \\ \midrule
Input             & 3.517                               & 1.984          & 2.484          & 0.797          \\
Encoder           & 3.639                               & 1.081          & 2.687          & 0.862          \\
FRN               & \textbf{3.655}                      & \textbf{0.946} & \textbf{2.797} & \textbf{0.889} \\ \bottomrule
\end{tabular}
}
\caption{Results of a variation from our proposed model on the `Real trace' test set.}
\label{tab:ablt}
\end{table}
With reference to Table~\ref{tab:ablt}, although the encoder alone achieved good performance, the predictor and joiner provides additional performance improvement. This result also implies that reusing the previous output is beneficial for this task.
\vspace{-0.3cm}
\section{Conclusions}
\label{sec:conclude}
\vspace{-0.15cm}
We proposed a deep recurrent model for blind packet loss concealment at 48~kHz. Despite the lack of loss trace information and being a frame-causal model, our method outperforms one of the informed models in the PLC challenge. The proposed FRN exhibits performance improvement over an offline model across the evaluation benchmarks by a significant margin. 
% With data augmentation employed, the model has become flexible to work with lower-band signals, which is essential for real-world applications.
\vfill\pagebreak

\bibliographystyle{IEEEbib}
\bibliography{IEEEabrv, refs}

\end{document}